\documentclass[doublecol]{epl2}

\usepackage{graphicx}
\usepackage{dcolumn}
\usepackage{bm}

\title{Exploring proteins multi-funnel energy landscape.}

\author{L. Cruzeiro}

\institute{CCMAR and FCT, University of Algarve, Campus de Gambelas,
8000-810 Faro, Portugal}

\pacs{87.15.Aa}{Theory and modeling; computer simulation}
\pacs{87.14.Ee}{Proteins} \pacs{87.15.He}{Dynamics and
conformational changes}

\abstract{An all-atom model of proteins is used to show that the
same sequence of amino acids can have many alternative structures,
that are very distant from, and that can be as stable as, the
corresponding native structure. Such alternative structures are not
easily rationalized as belonging to the native basin and indicate
instead that the free energy landscape of proteins is
multi-funnel-shaped and that Anfinsen's thermodynamic hypothesis
alone cannot explain protein folding. An alternative two-step
process for folding is proposed and its consistency with the
experimental evidence available is discussed.}

\begin{document}

\maketitle


\maketitle

\section{Introduction}
An outstanding question in Biology and Medicine, known as the
protein folding problem, is how a given sequence of amino acids, in
cells, most of the times assumes the native structure
\cite{lev68,anf73}. An important concept is that of the free energy
landscape and the current working hypothesis is that this landscape
is funnel-shaped \cite{bry,onu97,wol05,kar05} and that the native
structure corresponds to its global minimum
\cite{anf73,bry,onu97,wol05,kar05}. However, a question arises about
the consistency between the funnel hypothesis and the interactions
that stabilize protein structure. These interactions are reasonably
well represented by potentials such as these \cite{amber}:
%
%
%
\begin{eqnarray}
V & = & \sum_{\rm bonds} K_r (r - r_{eq})^2
                     + \sum_{\rm angles} K_\theta (\theta -
                     \theta_{eq})^2 +
\\ \nonumber
                    & + & \sum_{\rm dihedrals} {V_n \over 2}
                                       [1 + {\rm cos}(n\phi -
                                       \gamma)] +
                     \\ \nonumber
                     & + & \sum_{i<j} \left [ {A_{ij} \over R_{ij}^{12}} -
                                          {B_{ij} \over R_{ij}^6} +
                                          {q_iq_j \over \epsilon R_{ij}}
                                 \right ]
            \label{eqnff}
\end{eqnarray}
where bond stretching and bond bending (the first two sums) are
harmonic, rotations around a bond are described by a truncated
Fourier series (third sum) and nonbonded interactions are modelled
by the Lennard-Jones potential and Coulomb interactions due to the
partial charges on each atom (the last sum). A few systematic
studies of the shapes of the energy landscape of small polypeptides
and water clusters using these kind of potentials have been
attempted, which show both funnelled and multi-funnelled landscapes
\cite{bk97,lb98,wmw98,mw01}, with the local topography of the energy
landscape being related to the conformation of the molecule
 \cite{lb98}. Furthermore, a 4 $\mu$s study of the free energy landscape of a 16 amino
acid beta-hairpin led to three well defined non-native basins with
free energies comparable to that of the native basin \cite{kar04}.
On the other hand, the conformational space of proteins, albeit
small, remains too large to be probed in a systematic manner, even
with the most powerful computers.
Instead, here a cursory study of the energy landscape of four
proteins is made, taking their native basin as the reference.
\\

Using the nomenclature of the Protein Data Bank (PDB) \cite{pdb},
the four proteins are: 1QLX  (104 amino acids) \cite{1qlx}, 1I0S
(161 amino acids) \cite{1i0s}, 1AAP (56 amino acids) \cite{1aap} and
1IGD (61 amino acids) \cite{1igd}. These proteins have different
sizes, different biological origins and different functions. While
the first is a fragment of the human prion \cite{1qlx}, the second
is an oxireductase from archae \cite{1i0s}, the third is the
protease inhibitor domain of Alzheimer's amyloid $\beta$-protein
\cite{1aap} and the fourth is a immunoglobulin binding domain of
streptococcal protein G \cite{1igd}. The main criterion for their
selection was to have one representative of each of the four main
classes of proteins identified in the CATH hierarchical structural
classification scheme \cite{cath}: mainly $\alpha$ (1QLX), mainly
$\beta$ (1I0S), essentially structureless (1AAP) and $\alpha/\beta$
(1IGD). To probe the energy landscape of these four proteins, for
each one, three alternative structures were built by forcing it to
assume the fold, or part of the fold, of each of the other three, as
explained in detail below, and the stability of the resulting
structures was
compared with that of the corresponding native structure. \\

\begin{figure}
    \includegraphics[width=8.5cm]{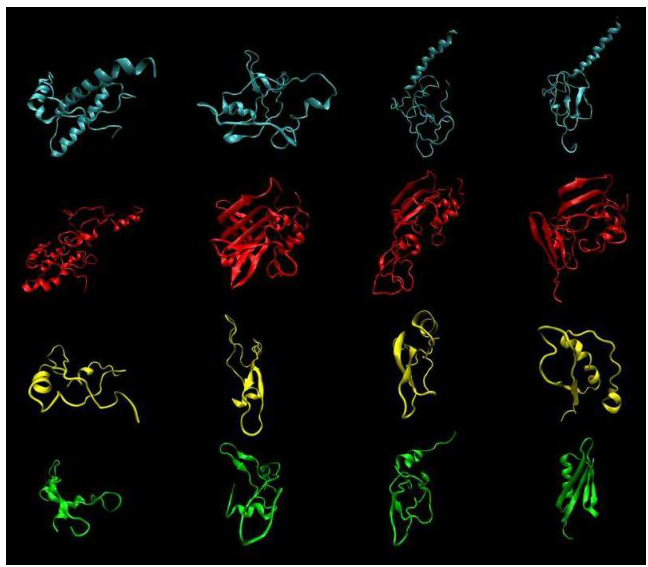}
\caption{(Colour online) Protein structures at the end of the 0.8 ns
equilibration period at 298~K. All proteins in the same row (with
same colour) have the same amino acid sequence. The four native
structures are displayed along the diagonal. The first row has the
structures for protein 1QLX (cyan), the second for 1I0S (red), the
third for 1AAP (yellow) and the fourth is for 1IGD (green). Along
each column, the non-native structures are obtained by imposing the
backbone fold, or part of the backbone fold, of the native structure
in that column on to the backbone of the other proteins. This figure
was made with VMD \cite{vmd}.} \label{fig1}
\end{figure}

The coordinates for the atoms in the native structures of the four
proteins selected were taken from the PDB \cite{pdb} and their
structures were energy minimized with the AMBER force field
\cite{amber}, to relieve any steric or otherwise strongly
unfavorable interactions. Alternative structures for each protein
were then built using the energy minimized native structures as
templates, by forcing the sequence of each protein to have the
backbone fold, or part of the backbone fold, of each of the other
three proteins. For example, the initial coordinates for the
structure in the first row, second column of figure \ref{fig1} were
obtained by imposing the backbone fold of the first 104 amino acids
of 1I0s onto the backbone of the 104 amino acids of 1QLX and the
initial coordinates for the second row, first column were obtained
by imposing the backbone fold of the 104 amino acids of 1QLX on to
the backbone of the first 104 amino acids of 1I0s. These and the
other alternative structures thus generated were first relaxed, in
order to eliminate all the steric interactions such a procedure
leads to, and, after relaxation, they were energy minimized
\cite{avail}. All 16 energy minimized structures were solvated in
water using the box option of the leap program of AMBER \cite{amber}
and the resulting systems were energy minimized, keeping the protein
fixed. Then the NAMD program \cite{namd}, with the AMBER force
field, was used to heat each of the systems to 298 K and to
equilibrate them at that temperature for 0.6 ns. A representative
statistical ensemble, at 298~K, for each of the 16 systems thus
constructed, was obtained by storing snapshots every picosecond from
a further 0.2 ns molecular dynamics (MD) run. It should be noted
that the smallest system in these simulations has 12234 atoms and
the largest has 33451 atoms.
\\

\section{Results}
Figure \ref{fig1}, which was made with the program VMD \cite{vmd},
shows the native folds of the four proteins and also the twelve
alternative folds, at the end of the MD sampling run. The data in
tables 1 and 2 was calculated from the same statistical sample. The
native structures for the four proteins are found along the diagonal
of figure \ref{fig1} and each row includes the native fold plus its
three alternative structures, all in the same colour. All proteins
in the same column were generated to have at least part of the fold
of the native structure in that column.
Inspection of figure \ref{fig1} shows that, even after heating and
equilibration at 298 K, the alternative structures retain most of
the backbone folds that were imposed on them initially, even if
these lead to very unnatural protein structures, particularly for
the amino acids sequences concerned.
\\

The average energies of all sixteen structures are displayed in
table 1, in which the data is organized in the same manner as in
figure \ref{fig1}. All systems in the same row of table 1 are
exactly the same, i.e., not only do they have the same protein and
ions but also the same number of water molecules, namely, the N
water molecules closest to the protein. The number N was chosen as
that for which the interaction energy between the protein and water
reached an average saturation value. For  the 1QLX and 1I0S proteins
N is 4000 and for the smaller proteins 1AAP and 1IGD N is 2000. In
each cell of table 1, the first number is the total energy of the
system constituted by the protein plus ions plus N water molecules
and is dominated by the water-water interactions. The second number
in each cell is the total energy of the protein, including the
intra-protein interactions (third number), the protein-ions
interactions (fourth number) and the protein-water interactions
(fifth number). Inspection of table 1 shows that some of the
alternative structures have equivalent, or even lower, potential
energies than the native structure. For example, imposing part of
$\beta$-fold of 1I0S on the naturally mainly $\alpha$ structure of
1QLX leads to a structure that has an average potential energy lower
than the native structure of 1QLX and imposing the essentially
disordered structure of the native fold of 1AAP on to the first 56
amino acids of 1I0S leads to a structure with an average energy
approximately equal to that of the native fold of 1I0S. Furthermore,
in the case of the protein 1AAP, all three alternative structures
have energies that are lower than its native fold and in the case of
1IGD, two of the alternative structures have lower energies.
\\
\begin{table*}
\caption{\bf Average Energies (kcal/mol)}
\begin{center}
{\footnotesize \bf
\begin{tabular}{|c|c|c|c|c|c|} \hline
Seq & mainly $\alpha$ &
mainly $\beta$ & disordered & $\alpha/\beta$
\\
 & Str 1QLX & Str 1I0S & Str 1AAP & Str 1IGD \\ \hline
 1 & -38773$^{a}$ $\pm$ 150 & -38067$^{a}$ $\pm$ 168 & -38650$^{a}$ $\pm$ 140 & -38224$^{a}$ $\pm$ 148
\\ Q
 &  -6428$^{b}$ $\pm$  65 &  -6709$^{b}$ $\pm$  73 &  -6333$^{b}$ $\pm$  78 &  -6242$^{b}$ $\pm$  88
\\ L
 &  -2619$^{c}$ $\pm$  50 &  -1854$^{c}$ $\pm$  47 &  -2573$^{c}$ $\pm$  56 &  -2531$^{c}$ $\pm$  58
\\ X
 &    -97$^{d}$ $\pm$  32 &   -363$^{d}$ $\pm$  57 &   -177$^{d}$ $\pm$  25 &   -138$^{d}$ $\pm$  41
\\
 &  -3712$^{e}$ $\pm$  74 &  -4492$^{e}$ $\pm$  76 &  -3584$^{e}$ $\pm$  80 &  -3573$^{e}$ $\pm$ 118
\\ \hline
 1 & -38404$^{a}$ $\pm$ 161 & -38796$^{a}$ $\pm$ 152 & -38173$^{a}$ $\pm$ 164 & -38089$^{a}$ $\pm$ 147
\\ I
 &  -7332$^{b}$ $\pm$  80 &  -7438$^{b}$ $\pm$  85 &  -7437$^{b}$ $\pm$  87 &  -7088$^{b}$ $\pm$  90
\\ 0
 &  -1534$^{c}$ $\pm$  67 &  -1948$^{c}$ $\pm$  72 &  -1562$^{c}$ $\pm$  58 &  -1533$^{c}$ $\pm$  80
\\ S
 &   -443$^{d}$ $\pm$  36 &   -220$^{d}$ $\pm$  42 &   -597$^{d}$ $\pm$  40 &      3$^{d}$ $\pm$  35
\\
 &  -5355$^{e}$ $\pm$  83 &  -5270$^{e}$ $\pm$ 113 &  -5278$^{e}$ $\pm$ 101 &  -5558$^{e}$ $\pm$ 110
\\ \hline
 1 & -18493$^{a}$ $\pm$ 128 & -18663$^{a}$ $\pm$ 113 & -18140$^{a}$ $\pm$ 135 & -18305$^{a}$ $\pm$ 137
\\ A
 &  -3334$^{b}$ $\pm$  71 &  -3452$^{b}$ $\pm$  51 &  -3078$^{b}$ $\pm$  54 &  -3413$^{b}$ $\pm$  52
\\ A
 &   -671$^{c}$ $\pm$  53 &   -535$^{c}$ $\pm$  40 &   -924$^{c}$ $\pm$  29 &   -622$^{c}$ $\pm$  40
\\ P
 &   -523$^{d}$ $\pm$  28 &   -775$^{d}$ $\pm$  42 &   -249$^{d}$ $\pm$  48 &   -815$^{d}$ $\pm$  35
\\
 &  -2140$^{e}$ $\pm$  93 &  -2141$^{e}$ $\pm$  51 &  -1904$^{e}$ $\pm$  57 &  -1975$^{e}$ $\pm$  51
\\ \hline
 1 & -18353$^{a}$ $\pm$ 128 & -18478$^{a}$ $\pm$ 103 & -18281$^{a}$ $\pm$ 122 & -18315$^{a}$ $\pm$ 109
\\ I
 &  -3252$^{b}$ $\pm$  58 &  -3299$^{b}$ $\pm$  53 &  -2943$^{b}$ $\pm$  55 &  -3060$^{b}$ $\pm$  58
\\ G
 &   -466$^{c}$ $\pm$  35 &   -323$^{c}$ $\pm$  37 &   -572$^{c}$ $\pm$  46 &   -598$^{c}$ $\pm$  38
\\ D
 &   -202$^{d}$ $\pm$  45 &   -591$^{d}$ $\pm$  24 &    -86$^{d}$ $\pm$  24 &     23$^{d}$ $\pm$  36
\\
 &  -2584$^{e}$ $\pm$  65 &  -2385$^{e}$ $\pm$  56 &  -2285$^{e}$ $\pm$  84 &  -2485$^{e}$ $\pm$  66
\\ \hline
\end{tabular}
}
\\
\end{center}
$^a$ Total energy of the system including protein, ions and the N
closest water molecules (see text)). $^b$ Total energy of the
protein, including $^c$ all the atom-atom interactions in the
protein plus $^d$ the ion-protein interactions and $^e$ the
protein-water interactions.
\end{table*}

In order to have an insight into the entropy associated with each
structure, the root mean square deviations (RMSD) per atom of each
of the structures with respect to its thermal equilibrium average
structure are presented in table 2 (only non-hydrogen backbone atoms
are used). The data indicate that the native structures fluctuate
less than the alternative structures in all cases, something that
reinforces the thermodynamic viability of the alternative structures
mentioned above.
\\
\begin{table*}
\caption{\bf Average fluctuations (\AA)}
\begin{center}
{\footnotesize \bf
\begin{tabular}{|l|c|c|c|c|c|}
\hline Seq & mainly $\alpha$ & mainly $\beta$ & disordered &
$\alpha/\beta$ \\
 & Str 1QLX & Str 1I0S & Str 1AAP & Str 1IGD \\ \hline
 1QLX &    0.85  $\pm$   0.06 &    1.26  $\pm$   0.17 &    1.26  $\pm$   0.22 &    1.16  $\pm$   0.16
\\ \hline
 1I0S &    1.15  $\pm$   0.15 &    0.89  $\pm$   0.09 &    0.98  $\pm$   0.12 &    1.08  $\pm$   0.16
\\ \hline
 1AAP &    1.23  $\pm$   0.13 &    1.20  $\pm$   0.13 &    0.84  $\pm$   0.07 &    1.08  $\pm$   0.15
\\ \hline
 1IGD &    1.17  $\pm$   0.18 &    1.24  $\pm$   0.20 &    1.16  $\pm$   0.19 &    0.92  $\pm$   0.12
\\ \hline
\end{tabular}
}
\end{center}
\end{table*}

As a further qualitative measure of the relative structural
stability of the structures, all were heated from the equilibrium
value of 298 K to a final value of 698 K, at the rate of 2 K per ps.
Figure 2 shows the variation of the RMSD per atom of each structure
during this extra heating procedure, with respect to the
corresponding initial structure. The plots are organized per
sequence, in the same order as in the previous figure and tables,
and, in each plot, the solid line is for the native structure. A
general trend is that native structures take longer to deviate from
the initial structure, something that is in agreement with the fact
that they fluctuate less while at thermal equilibrium at 298~K.
However, some of the alternative structures have a very similar
behaviour to their corresponding native structure. Also, other
general trends are that all structures show the same average
structural stability until 50 ps (when the temperature has increased
to 398 K) and that the greatest divergence takes place at 100 ps for
the smaller proteins 1AAP and 1IGD (when the temperature has
increased to 498 K) and at 150 ps, when the temperature is 598 K,
for the larger proteins 1QLX and 1I0S. Thus, this qualitative
measure of activation energies for conformational changes shows
that, given a certain amino acid sequence, it is possible to find
structures that are very different from the native and yet have a
similar structural stability.
\\
\begin{figure}
    \includegraphics[width=8.4cm]{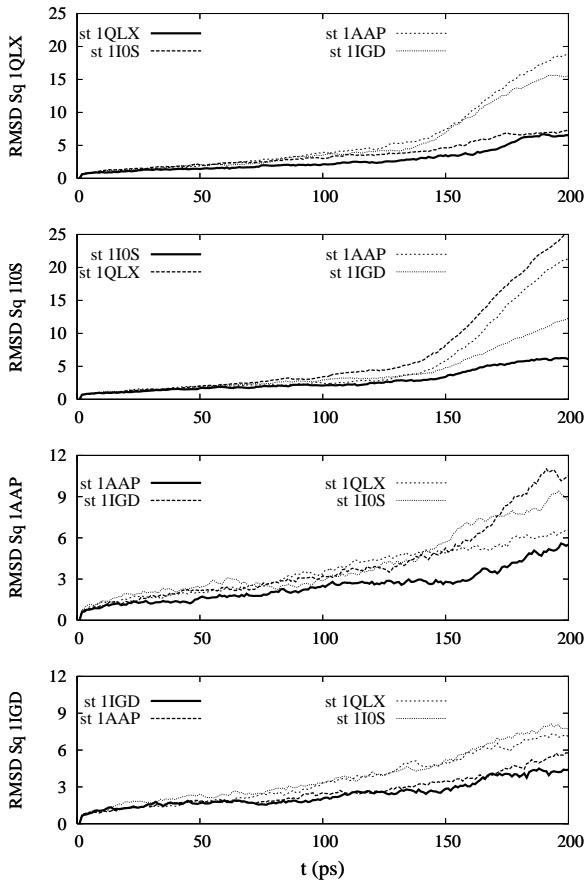}
\caption{Variation of the RMS deviation of each snapshot with
respect to the corresponding initial structure. The temperature
increased throughout the simulation at the rate of 2 K per ps. Each
plot is for a different protein and the order from top to bottom is
as in Figure 1. In each plot, the solid lines are for the native
structures and the remaining three lines are for the proteins
indicated by the keys. The values are in \AA.} \label{fig2}
\end{figure}

\section{Discussion}
The native plus the three alternative structures studied here
provide a mere glimpse into the conformational space of each of the
four proteins selected. They were built by taking proteins whose
sequences, in cells, lead them a particular structural class and by
making them assume structures of other, very different, classes of
proteins. This protocol was applied to proteins of different sizes
and without any regard as to how best to fit the alternative folds
to each particular protein, i.e., one could argue that the
alternative structures studied here constitute examples of the most
unnatural alternative conformations that we can obtain for the four
proteins selected. Nevertheless, we find that some of them have
energies that are at least comparable, and in some cases, even more
favorable, than the native structures. Furthermore, the alternative
structures generated are sufficiently separated from the native
configuration to make transitions to it improbable in normal
conditions of temperature and pressure. One limitation of the
simulations reported here is their time length, which is short when
compared with experimental times for conformational changes. While
no transitions from the non-native folds to the native folds were
observed, it cannot be ruled out that such transitions might occur
at longer times. It should be noted, however, that the fluctuations
observed for the 16 structures are between $0.85$ and $1.26$~\AA,
approximately the same as in the structures determined in NMR
measurements \cite{rmsd}. This is due to a coupling of the thermal
bath that is much stronger in the simulations and one consequence is
that the computational time for conformational changes to take place
is shorter. Also, an indication of the relative size of the energy
barriers for conformational changes can be obtained by heating the
systems and, in this respect, figure \ref{fig2} shows that, although
the barriers for the native structures are higher, they do not seem
to be significantly different from those of the non-native
structures. Thus, while the results presented here cannot be said to
prove it they do suggest that
the free energy landscape of proteins has the shape of a
multi-funnel, in which each funnel is associated with an average
structure that can be very different from the native, and yet be as
thermodynamically viable as the native structure.
\\


Although the current theoretical framework for protein folding is
based on a funnel-shaped free energy landscape
\cite{bry,onu97,wol05,kar05}, experimental evidence for a
multi-funnel free energy landscape in the case of proteins was first
obtained in 1968 by Levinthal who found two forms of an alkaline
phosphatase at 317~K, one active and the other inactive, synthesized
at different temperatures, in mutants of E. Coli \cite{lev68}. More
recently, other cases have been found of proteins that can assume
more than one structure in the same thermodynamical conditions
\cite{pru82,pru06,bsa92,sja98,serpin}.
While a funnel free energy landscape has difficulty in explaining
why protein misfolding happens, a multi-funnel free energy landscape
can readily rationalize it as a case in which a non-native funnel
was selected. On the other hand, a multi-funnel free energy
landscape cannot explain protein folding just by a principle of free
energy minimization. Indeed, in a multi-funnel, the difficulty in
determining the native structure from a given amino acid sequence is
not just due to the size of the conformational space and the lack of
computer power, but, more essentially, to the fact that the native
structure is not a well-defined global free energy minimum.
\\

Given the experimental evidence, and the results presented here and
elsewhere \cite{kar04}, it is worthwhile to start thinking about how
proteins can fold to a well-defined average structure in a
multi-funnel free energy landscape. One possibility is that protein
folding involves two steps, a first step in which a specific funnel
is selected (most of the times that funnel being the native funnel)
and a second step in which the structure relaxes as its energy is
minimized within that funnel \cite{jbp01}. The first step is a
kinetic mechanism for which there is direct experimental evidence in
a few cases \cite{bsa92,sja98,serpin} and that was already proposed
by Levinthal who suggested that there are specific pathways for
folding \cite{lev68}. Considering that such pathways can be
characterized by intermediates one can say that the experimental
evidence for a kinetic mechanism is indeed substantial
\cite{englander,rad07} and may even include proteins that apparently
follow a two-state process \cite{rad07,roder}. The second step is an
energy minimization mechanism, as first proposed by Anfinsen
\cite{anf73} and incorporated in the funnel models
\cite{bry,onu97,wol05,kar05}. Within this two-step picture of
folding, proteins denature reversibly as long as heating does not
make them diffuse away from the native funnel and denature
irreversibly otherwise. Chemical unfolding, on the other hand,
cannot be described within a single free energy landscape picture
because as the denaturant concentration varies, so does the system
and consequently the associated free energy landscape.
\\

It is curious to note that experimental evidence also points to the
existence of two steps in protein folding, one in which a compact
structure forms, in the dead time of the experiments \cite{roder}
and another, which takes much longer, from microseconds to
milliseconds or more, at the end of which proteins become active.
The suggestion here is that the first step is related to the
selection of the funnel, while the second step is due to the
relaxation down the funnel selected. In this two-step picture, the
second step is the rate-limiting step, i.e. the rates of folding are
dependent on this (slow) relaxation down the funnel selected, but
the definition of the structure is accomplished in the first step,
when a particular funnel is selected. Thus, within a multi-funnel
free energy landscape, to understand how a given amino acid sequence
leads to a specific three dimensional structure one must understand
the kinetic mechanism by which a specific funnel is selected.
\\

\acknowledgments This work was funded by the Foundation for Science
and Technology (FCT, Portugal) and by POCI 2010 and the European
Community fund, FEDER. Most of the computer simulations were
performed at the Laboratory for Advanced Computing (LCA), University
of Coimbra, Portugal.


\begin{thebibliography}{0}
\bibitem{avail}
All 16 minimum energy structures, four of which have the native fold
and 12 of which are alternative, hybrid, structures are available,
in PDB format, upon request.
\end{thebibliography}
\end{document}